\documentstyle[11pt,newpasp,twoside,psfig]{article}
\def\edcomment#1{\iffalse\marginpar{\raggedright\sl#1\/}\else\relax\fi}
\reversemarginpar

\def\cm{{\rm\thinspace cm}}

\def\erg{{\rm\thinspace erg}}

\def\km{{\rm\thinspace km}}

\def\s{{\rm\thinspace s}}
\def\erg{{\rm\thinspace erg}}

\def\ergpcmsqps{\hbox{$\erg\cm^{-2}\s^{-1}\,$}}

\def\ergps{\hbox{$\erg\s^{-1}\,$}}

\def\pcmsq{\hbox{$\cm^{-2}\,$}}
\def\kmps{\hbox{$\km\s^{-1}\,$}}
\def\feka{$F_{\rm K\alpha} \;$}
\def\fexxv{Fe~{\sc xxv}}
\def\fexxvi{Fe~{\sc xxvi}}
\def\feka{$F_{\rm K\alpha} \;$}

\def\approxlt{\mathrel{\spose{\lower 3pt\hbox{$\sim$}} 
	\raise 2.0pt\hbox{$<$}}}
\def\approxgt{\mathrel{\spose{\lower 3pt\hbox{$\sim$}}
        \raise 2.0pt\hbox{$>$}}}

\def\grs1915{GRS~1915+105}
\def\chandra{{\it Chandra }}
\def\rxte{{\it RXTE }}
\def\asca{{\it ASCA }}

\begin{document}
\title{Chandra$-$ASCA$-$RXTE observations of the micro-quasar \grs1915}
 \author{Julia C. Lee,  Norbert S. Schulz}
\affil{MIT Center for Space Research, 77 Massachusetts Ave. NE80, Cambridge, MA. 02139 }
\author{Christopher S. Reynolds}
\affil{JILA - U. of Colorado, Campus Box 440, Boulder CO 80309 }
\author{Andrew C. Fabian}
\affil{Institute of Astronomy, Madingley Rd., Cambridge CB2 0HA  U.K.}
\author{Eric G. Blackman}
\affil{Dept. of Physics \& Astronomy, University of Rochester, Rochester, NY 14627 }

\begin{abstract}
 A {\it Chandra} AO1 30ks HETGS observation of the X-ray transient
micro-quasar GRS~1915+105 reveals absorption edges and faint line
emission over the HETG energy range.  We find from a preliminary
analysis evidence for prominent neutral K edges associated with
iron, silicon, magnesium, and tentatively sulphur.  The column densities assuming 
solar abundances are consistent with 
$\sim \rm few \times 10^{22} \, cm^{-2}$ in excess of the Galactic value,
and may point to surrounding cold material associated with \grs1915.  
Neutral \feka emission, and ionized absorption from Fe~{\sc xxv} and  Fe~{\sc xxvi}
are resolved. We limit our discussion to the \chandra results.
\end{abstract}

\section{Introduction}
The X-ray transient GRS~1915+105 is an extremely energetic 
object in our Galaxy.  Since its discovery by {\it Granat/WATCH} in 1992
(Castro-Tirado et al. 1992), it has shown repeated flares separated by
short periods of quiescence.  GRS~1915+105 has been studied extensively in
multiple wavebands which include radio, infra-red and X-rays (e.g. see
review by Zhang et al. 1997).  In the X-ray band, its overall variability
is extremely complex (possessing both regular and chaotic patterns) and
challenges any current theoretical model.  In its flaring state, the
total (isotropic) luminosity can exceed $10^{39}\ergps$, making it one of
the most energetic objects known in the Galaxy.  Since this exceeds the
Eddington limit for a neutron star by almost an order of magnitude, it
seems probable that this system contains a black hole. Given its distance
of $\sim$~12.5\,kpc (Mirabel \& Rodr\'{\i}guez 1994) determined from
the 21~cm absorption of atomic hydrogen along the line of sight,  and a visual extinction of 
$A_V\sim 27$, a direct optical determination of the mass function of this 
system is not possible -- partly for this reason, the nature of the companion
(i.e. LMXRB -- e.g. Castro-Tirado et al. 1996 versus HMXRB -- e.g. Chaty et al.
1996) is still debated.  It is highly variable, with a luminosity which
varies between a few $\times 10^{37}-10^{39}
\ergps$, or in excess of this during the flaring state.  
One of the most remarkable aspects of this system is that X-ray flares are
followed by superluminal ejection events which are observed in the radio band (Mirabel \& Rodr\'{\i}guez 1994) --
the inferred velocity of the jets (or more likely blobs) is $v\approx 0.9c$
(i.e. a Lorentz factor of 3), at an inclination $\approx 60-70^\circ$ to the 
line of sight.  The Galactic binaries GRO~1655--40
and XTE~J1748--288 display similar
behavior.  In this respect, these sources qualitatively resemble
radio-loud active galactic nuclei (AGN) and are often called
`micro-quasars'.

\section{Observation}
\grs1915 was observed with the {\it Chandra} High Energy Transmission
Grating (HETGS; Canizares et al., in preparation) on 2000 April 24 
with a total integration time of $\sim$~31.4~ks (Fig.~1).
Simultaneous \rxte observations were performed from 2000 April 24$-$25
with both the Proportional Counter Array (PCA) and
High-Energy X-ray Timing Experiment (HEXTE) instruments.
An \asca target of opportunity (TOO) between 2000 April 19$-$25
also coincides with our \chandra observation.  We concentrate
only on the results of the \chandra observations for this proceeding.

The \chandra HETG is made up of the Medium Energy Grating (MEG) with
a 2.5-31~\AA\, (0.4-5~keV) bandpass, and High Energy Grating (HEG) 
with 1.2$-$14~\AA\,(0.9-10~keV) bandpass.  The peak resolving power (E/$\Delta$E)
respectively for the MEG and HEG are 520 at 1~keV, and 1000 at 1~keV.
The resolution of the higher orders improves by a factor of $n$ for the $n$th
order, but the spectral bandpass and efficiency are reduced accordingly.

Due to severe telemetry and photon pileup problems imposed by the large count rate
of \grs1915, the observation was performed using the {\it graded} telemetry
mode, and a subarray which reduced CCD frametime by a factor of $\sim 2$.
Even though the source was at a relatively low state ($F_{\rm X} = 9 \times 10^{-9}
\ergpcmsqps$ which translates to $L_{\rm X} = 2.14 \times 10^{38} \ergps$; Fig.~1) during
the epoch of our observation, we were nevertheless hampered by severe pileup
in the HETG 1st orders.  Because we were prepared for this eventuality, a
`spatial window' was used to block out the 0th order image (this would be
completely piled-up even during periods of low flux for the energetics of \grs1915.)
However, because the 0th order position is crucial for defining an accurate wavelength 
scale, we calculate its position for our observation 
by fitting the intersection of the MEG and HEG dispersed images
with the 0th order readout trace.  The accuracy of such a technique
for the determination of the 0th order position is $\sim$ 0.2$-$0.3 pixels, which
translates respectively to a wavelength accuracy of 0.002~\AA \, and 0.004~\AA, 
for the HEG and MEG 1st order. 

\begin{figure*}
\begin{center}
\psfig{file=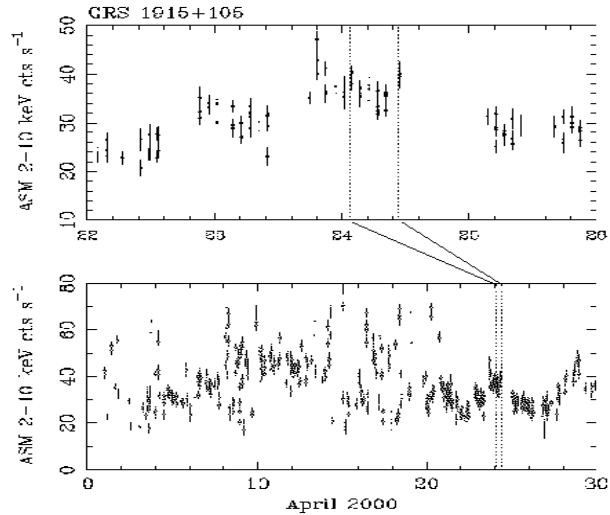,angle=270,width=14.5truecm,height=8.5truecm}
\caption[h]{The \rxte All-Sky Monitor (ASM) light curve of \grs1915 at the epoch of the \chandra-\asca-\rxte observations.  (The bottom light curve shows the behavior of \grs1915 during the month of August 2000 for comparison.) Dotted vertical lines denote the period of the 31~ks \chandra observation.  For comparison, 1~Crab $\approx$ 75~ASM ct\,$\rm s^{-1}$ $\approx$~$2\times 10^{-8}$ \ergpcmsqps.}
\end{center}
\end{figure*}

\begin{table}
\begin{center}
\begin{tabular}{lccc}
\multicolumn{4}{c}{\sc Prominent Spectral Features seen in the Time-averaged data} \\
\hline
\hline
\multicolumn{4}{c}{\sc Neutral Features} \\
{\em \rm Species} & {\em \rm $\lambda_{mea}$ (\AA)} & {$^a \Delta \lambda$ ($\kmps$)} & {$^b \rm N_H (10 ^{22} \pcmsq$)} \\
\hline
Mg K edge & 9.52 &  - & 2.7\\
Si K edge & 6.74 & - & 3.5 \\
S K edge& 5.02 &  - & 2.3 \\
Fe K edge& 1.74 & -  &  10.0\\
Fe K$\alpha$ emission & 1.94 & 1000 &-  \\
\hline
\multicolumn{4}{c}{\sc Ionized Features} \\
\hline
Fe~{\sc xxv} absorption & 1.86 & 600& - \\
Fe~{\sc xxvi} absorption & 1.78 & 1500 &  - \\
\end{tabular}
\end{center}
\caption{$^a$ Velocity width (instrumental response removed). $^b$ Hydrogen column in
excess of the Galactic ($1.76 \times 10^{22} \pcmsq$) value,
assuming solar abundances.}
\end{table}

\begin{figure*}
\psfig{figure=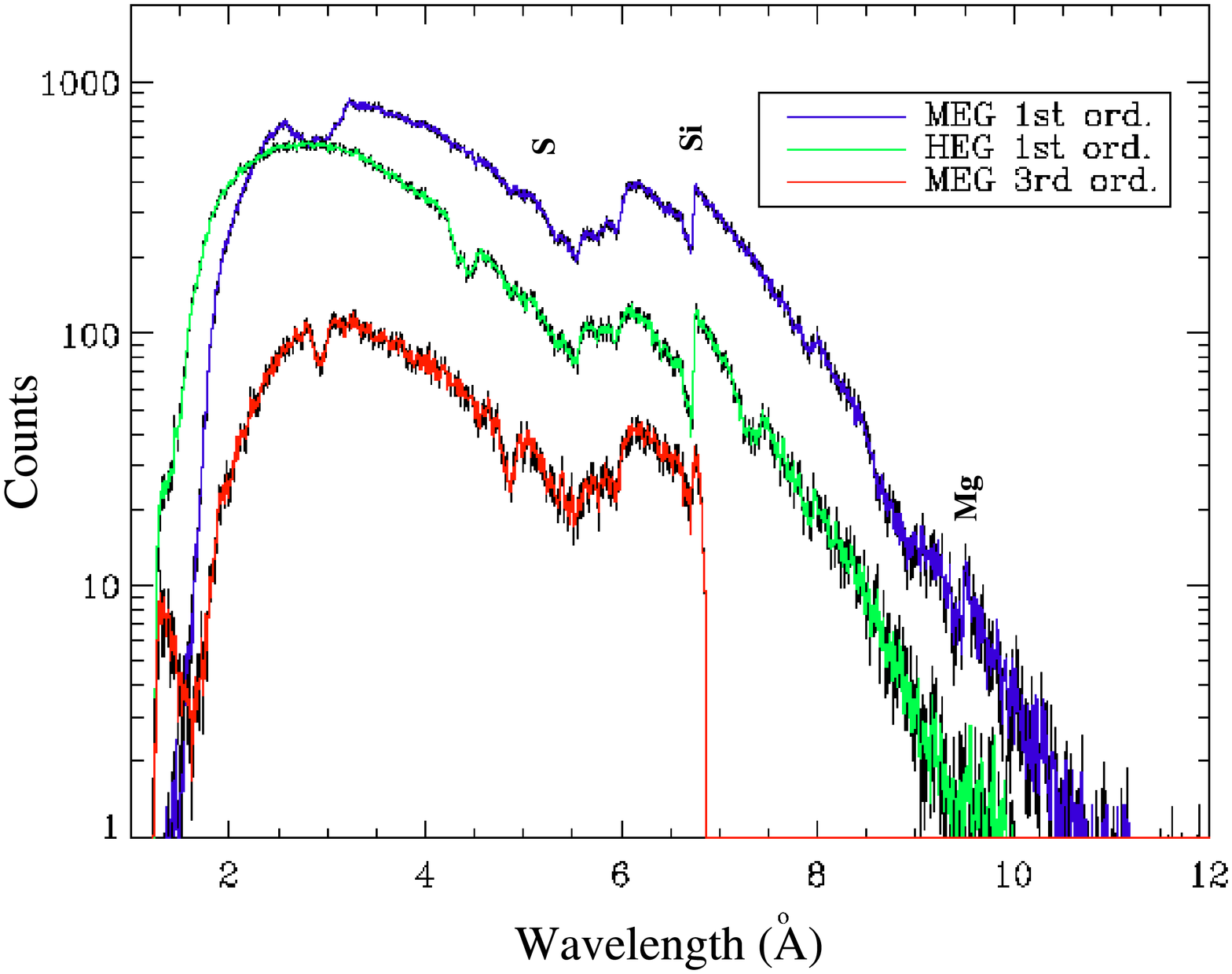,angle=270,width=0.95\textwidth,height=0.4\textheight}
\psfig{figure=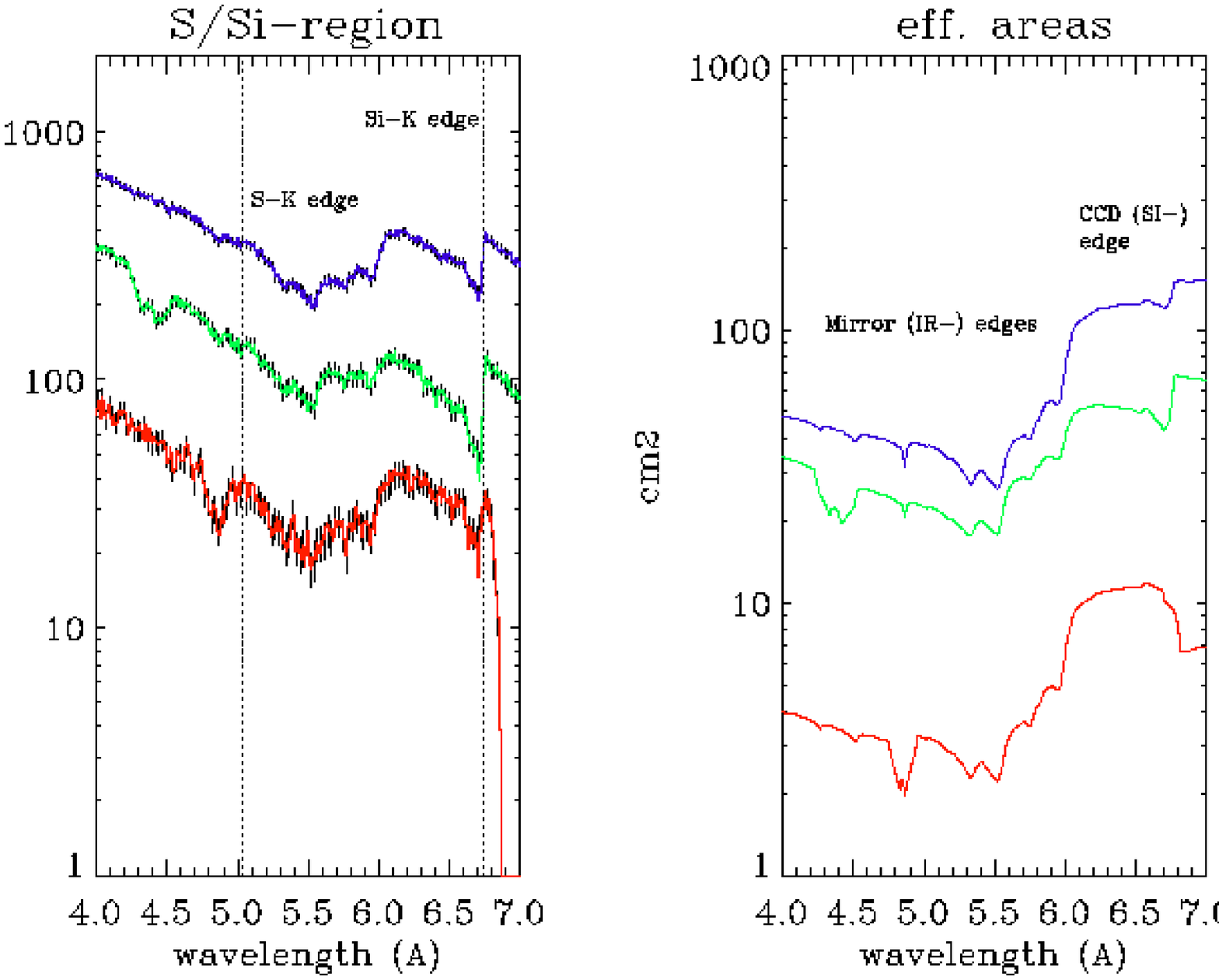,width=0.95\textwidth,height=0.5\textheight,angle=0 }
\caption[h]{ (Top) \grs1915 count spectrum from top spectrum down : MEG 1st order, HEG 1st order, MEG 3rd order
with three of the 4 neutral edges marked.
(Bottom) The S-Si region is compared with the effective area curves for the detectors
of interest (as in above plot scheme).  Note the prominence
of the Si edge in the MEG 3rd order spectrum where there is very little contribution from the detector edge.}
\label{grs1915-edge}
\end{figure*}

\begin{figure*}[t]
\psfig{figure=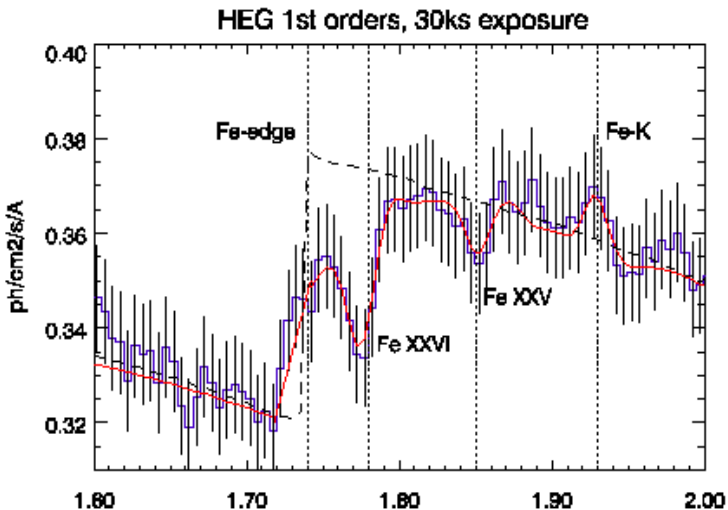,angle=0,width=0.95\textwidth,height=0.4\textheight}
\caption{The iron region in the HEG band.  Dotted lines represent the best fit continuum
model. Neutral and ionized features are superposed on the continuum. The complete
model (dashed lines which trace the spectrum) includes the continuum and absorption/emission features.}
\label{grs1915-feregion}
\end{figure*}

\section{Cold and Ionized Material}
We find evidence for prominent absorption edges due to iron, silicon, magnesium,
and tentatively sulphur at the expected wavelengths for the neutral K-edges
(Table~1, Figs.~2~and~3).  The column in excess
of Galactic ($1.76 \times 10^{22} \pcmsq$ determined from 21~cm absorption of atomic
hydrogen along the line of sight) is calculated using the value for the K-shell
photoionization cross section of the relevant species (Daltabuit \& Cox 1972),
and solar abundances. (We have used these abundances for comparison and acknowledge
that there is no reason to prefer these abundances over interstellar abundances.)
 The validity of the Si edge is checked against the 
contribution from the \chandra CCDs - we find an upper limit of $\sim$20\% contamination
from the detector to the depth of the edge.  The validity that the Si edge is intrinsic
to the source is reinforced in the MEG 3rd order spectrum (Fig.~2 bottom), where the edge remains
prominent in the data, but for which there is a minute contribution from the detector.
(The spectra for the relevant orders are the combined plus and minus sides;  in the case
of the MEG 3rd order, the Si edge seen in our spectrum for one of the sides fall on
the backside (S1) chip where there is no/little contribution from the  detector Si edge.)

In addition to neutral edges, neutral (Fe K$\alpha$) emission, and ionized iron (\fexxv, \fexxvi)
resonant absorption features are superposed on the continuum (Fig.~3), at the expected wavelengths.
The velocity widths and line wavelengths of these features are detailed in Table~1.
(The resolution of the \chandra HEG is $\sim 1600 \kmps$ at the iron energies.  The values
quoted in Table~1 already accounts for this.)

A first order approximation to the continuum is well fit by a simple Galactic absorbed cutoff power law
of $\Gamma \sim 2$, and edges.

\section{Summary}
While analysis is still at its preliminary stages, we find definite evidence 
for $N_H$ in excess of that expected from the Galactic column.
This can be an abundance effect (i.e. interstellar abundances), or also
likely cold material associated with \grs1915.  If the latter is true,
our \chandra results would confirm reports from near-IR VLT observations
of cold material associated with the source.
Mirabel et al. (1996) suggested that heated dust is associated 
with the surroundings of \grs1915.  Additionally, Mart\'{\i} et al. (2000) have
interpreted observed H{\sc i} P-Cygni profiles in the near-IR which turn into blue emission 
wings when the system is in outburst, to be due to material which surrounds
\grs1915 that is blown out during an X-ray outburst.   Our findings with
\chandra for neutral spectral features are consistent with associated cold
material in the environment of \grs1915.  The observed Doppler broadened iron
emission and absorption features are likely to 
be from the disk or corona, or even possibly a wind.

\end{document}